\documentclass[twocolumn,preprintnumbers,amsmath,amssymb]{revtex4}

\usepackage{epsfig,graphicx,dcolumn,bm,times}

\begin{document}

\title{A simple reaction-diffusion population model on scale-free networks}

\author{An-Cai Wu,$^{1,2}$ Xin-Jian Xu,$^{2,3}$ Jos\'{e} F. F. Mendes,$^{2,}$\footnote{Electronic address: jfmendes@ua.pt} and Ying-Hai Wang$^{1,}$\footnote{Electronic address: yhwang@lzu.edu.cn}}

\affiliation{
$^{1}$Institute of Theoretical Physics, Lanzhou University, Lanzhou Gansu 730000, China\\
$^{2}$Departamento de F\'{i}sica da Universidade de Aveiro, 3810-193 Aveiro, Portugal\\
$^{3}$Department of Mathematics, College of Science, Shanghai
University, Shanghai 200444, China}

\date{\today}

\begin{abstract}
We study a simple reaction-diffusion population model (proposed by
A. Windus and H. J. Jensen, J. Phys. A: Math. Theor. \textbf{40},
2287 (2007)) on scale-free networks. In the case of fully random
diffusion, the network topology does not affect the critical death
rate, whereas the heterogenous connectivity makes the steady
population density and the critical population density small. In the
case of modified diffusion, the critical death rate and the steady
population density are higher, at the meanwhile, the critical
population density is lower, which is good for survival of species.
The results are obtained with a mean-field framework and confirmed
by computer simulations.
\end{abstract}


\maketitle Recently, Windus and Jensen \cite{jpa07,tpb07} introduced
a model for population on lattices with diffusion and birth/death
according to $2A\rightarrow 3A$ and $A\rightarrow\phi$ for an
individual $A$ . They found that the model displays a phase
transition from an active to an absorbing state which is continuous
in $1+1$ dimensions and of first-order in higher dimensions
\cite{jpa07}. They also investigated the importance of fluctuations
and that of the population density, particularly with respect to
Allee effects in regular lattices \cite{tpb07}. It was found that
there exists a critical population density below which the
probability of extinction is greatly increased, and the probability
of survival for small populations can be increased by a reduction in
the size of the habitat \cite{tpb07}.

In the study of complex networks \cite{Albert}, an important issue
is to investigate the effect of their complex topological features
on dynamical processes \cite{review07}, such as the spread of
infectious diseases \cite{pv01a} and the reaction-diffusion (RD)
process \cite{GA04}. For most real networks, the connectivity
distribution has power-law tails $P(k)\sim k^{- \gamma}$, namely, a
characteristic value for the degrees is absent, hence the scale-free
(SF) property. In this Brief Report, we shall study the simple RD
population model \cite{jpa07} on SF networks.

In an arbitrary finite network which consists of nodes
$i=1,\ldots,N$ and links connecting them. Each node is either
occupied by a single individual (1) or empty (0). We randomly choose
a node. If it is occupied, the individual dies with probability
$p_{\rm d}$, leaving the node empty. If the individual does not die,
a nearest neighbor-node is randomly chosen. If the neighboring node
is empty, the particle moves there; otherwise, the individual
reproduces with probability $p_{\rm b}$ producing a new individual
on another randomly selected neighboring node, conditional on that
node being empty. A time step is defined as the number of network
nodes $N$. In homogeneous networks (such as regular lattices and
Erd{\"o}s-R{\'e}nyi (ER) random networks \cite{erdos59}), the
mean-field (MF) equation for the density of active nodes $\rho(t)$
is given by \cite{tpb07}
\begin{equation}
\frac{d \rho(t)}{d t} = -p_{\rm d}\rho(t) + p_{\rm b}(1-p_{\rm
d})\rho^{2}(t)(1-\rho(t)), \label{rateeq00}
\end{equation}
which has three stationary states
\begin{equation}
\bar\rho_0 = 0, \quad
\bar\rho_\pm=\frac{1}{2}\left(1\pm\sqrt{1-\frac{4p_{\rm d}}{p_{\rm
b}(1-p_{\rm d})}}\right). \label{steady states}
\end{equation}
For $4p_{\rm d} > p_{\rm b}(1-p_{\rm d})$, $\bar\rho_0$ is the only
real stationary state, and one can obtain a critical death rate
$p_{{\rm d}_{\rm c}} =p_{\rm b}/(4+p_{\rm b})$ which separates the
active phase representing survival and the absorbing state of
extinction. Simple analysis shows that $\bar\rho_+$ and $\bar\rho_0$
are stable stationary states, whereas $\bar\rho_-$ is unstable and
therefore represents a critical density $\rho_{\rm c}$ below which
extinction will occur in all cases. Thus, for $p_{\rm d} < p_{{\rm
d}_{\rm c}}$, one can write \cite{tpb07}
\begin{equation}
\rho(t) \rightarrow \left\{
\begin{array}{cl}
0 & \text{if} \quad \rho(t) < \rho_{\rm c}, \\
\bar\rho_+ & \text{if} \quad \rho(t) > \rho_{\rm c},
\end{array} \right.
\quad \text{as} \quad t \rightarrow \infty.
\end{equation}
At $p_{\rm d} = p_{{\rm d}_{\rm c}}$, the stationary density jumps
from $1/2$ to $0$, resulting in a first-order phase transition.

In order to study analytically this process on SF networks in which
the degree distribution has the form $P(k)\sim k^{- \gamma}$ and
nodes show large degree fluctuations, we are forced to consider the
partial densities $\rho_k(t)$, representing the density of
individuals in nodes of degree $k$ at time $t$ \cite{pv01a}. To
obtain a rate equation for $\rho_k(t)$ , we use a microscopical
approach which has been applied in diffusion-annihilation
\cite{da05} and multicomponent RD processes on SF networks
\cite{mrd06}. Let $n_i(t)$ be a dichotomous random variable taking
values $0$ or $1$ whenever node $i$ is empty or occupied by an
individual $A$, respectively. Using this formulation, the state of
the system at time $t$ is completely defined by the state vector
${\bf n}(t)=\{n_1(t),n_2(t),\cdots,n_N(t)\}$. The evolution of ${\bf
n}(t)$ after a time step can be expressed as
\begin{equation}
n_i(t+1)=n_i(t)\eta+[1-n_i(t)]\xi, \label{evolution}
\end{equation}
where $\eta$ and $\xi$ are dichotomous random variables, taking
values of $0$ or $1$ with certain probabilities $p$ and $1-p$,
respectively,

\begin{equation}
\eta= \left\{
\begin{array}{cl}
0; & \displaystyle{p=  p_{\rm d}+(1-p_{\rm d})
\left[1-\frac{1}{k_i}\sum_j a_{ij} n_j(t) \right]}, \\[0.5cm]
1; & \displaystyle{1-p},
\end{array}
\right.
\end{equation}
\begin{equation}
\xi= \left\{
\begin{array}{cl}
1; & \displaystyle p=   \displaystyle{\sum_j \frac{(1-p_{\rm
d})a_{ij} n_j(t)}{k_j} \left[1+\frac{p_{\rm b}}{k_j}\sum_l
a_{jl} n_l(t) \right]},  \\[0.5cm]
0; & \displaystyle{1-p}
\end{array}
\right.
\end{equation}
Obviously, $p$ is the probability that an occupied node $i$ becomes
empty. If node $i$ is empty, there are two factors causing it
occupied. One is that its survival neighbors will move to $i$ with
probability $\sum_j \frac{(1-p_{\rm d}) a_{ij} n_j(t)}{k_j}$ and the
other is that its survival neighbor $j$ selecting an occupied
neighbor $l$ (with the probability $\frac{1}{k_j}\sum_l a_{jl}
n_l(t))$) reproduces a new individual on $i$, then we have the term
$\sum_j \frac{(1-p_{\rm d}) a_{ij} n_j(t)}{k_j} \frac{p_{\rm
b}}{k_j}\sum_l a_{jl} n_l(t))$. Taking the average of
Eq.~(\ref{evolution}), we obtain
\begin{widetext}
\begin{eqnarray}
\langle n_i(t+1) | {\bf n}(t)\rangle &=& n_i(t)(1-p_{\rm
d})\frac{1}{k_i}\sum_j a_{ij} n_j(t)+(1-n_i(t))\left[\sum_j
\frac{(1-p_{\rm d}) a_{ij} n_j(t)}{k_j} (1+\frac{p_{\rm
b}}{k_j}\sum_l a_{jl} n_l(t))\right], \label{evolution_averaged}
\end{eqnarray}
\end{widetext}
which describes the average evolution of the system, conditioned to
the knowledge of its state at the previous time step. In the MF
approximation, $\langle n_i(t)n_j(t)\rangle \equiv \langle n_i(t)
\rangle \langle n_j(t) \rangle$ and $\langle
n_i(t)n_j(t)n_l(t)\rangle \equiv \langle n_i(t) \rangle \langle
n_j(t) \rangle \langle n_l(t) \rangle$. Thus, after multiplying
Eq.~(\ref{evolution_averaged}) by the probability to find the system
at state ${\bf n}$ at time $t$ and summing for all possible
configurations, we obtain
\begin{widetext}
\begin{eqnarray}
\rho_i(t+1) & \equiv & \langle n_i(t+1) \rangle =
\rho_i(t)\frac{(1-p_{\rm d})}{k_i}\sum_j a_{ij}
\rho_j(t)+(1-\rho_i(t))\left[\sum_j \frac{(1-p_{\rm d}) a_{ij}
\rho_j(t)}{k_j} (1+\frac{p_{\rm b}}{k_j}\sum_l a_{jl}
\rho_l(t))\right]. \label{evolution02}
\end{eqnarray}
\end{widetext}
We assume that nodes with the same degree are statistically
equivalent, i.e.,
\begin{equation}
\rho_i(t) \equiv \rho_k(t) \quad \forall i \in \mathcal{V}(k),
\end{equation}
and have
\begin{equation}
\sum_j a_{ij}=\sum_{k'}\sum_{j \in
\mathcal{V}(k')}a_{ij}=\sum_{k'}kP(k'|k) \quad \forall i \in
\mathcal{V}(k),
\end{equation}
where $\mathcal{V}(k)$ is the set of nodes of degree $k$.

We split the sum with index $j$ into two sums over $k'$ and
$\mathcal{V}(k')$, respectively. The double sum over $a_{ij}$ is
related to the conditional probability $P(k'|k)$ that a node of
given degree $k$ has a neighbor which has degree $k'$. In present
work, we restrict ourselves to the case of uncorrelated networks in
the following, in which the conditional probability takes the simple
form $P(k' | k) = k' P(k') / \langle k\rangle$. Thus, from
Eq.~(\ref{evolution02}) and after some formal manipulations, we
obtain
\begin{widetext}
\begin{equation}
\rho_k(t+1)=\rho_k(t)(1-p_{\rm d})\Theta(\rho(t))+\frac{k}{\langle
k\rangle} (1-p_{\rm d})(1-\rho_k(t))\rho(t)[1+p_{\rm
b}\Theta(\rho(t))],
   \label{evolution04}
\end{equation}
\end{widetext}
where $ \rho(t)$ is the total density of active individuals and
$\Theta(\rho(t))$ is the probability  that any given link points to
an occupied node
\begin{equation}
\Theta(\rho(t)) = \frac{1}{\langle k\rangle} \sum_k k P(k)
\rho_k(t). \label{eq:5}
\end{equation}
From Eq.~(\ref{evolution04}), we can obtain the following rate
equation
\begin{widetext}
\begin{equation}
\frac{d \rho_k(t)}{d t}=-\rho_k(t)+\rho_k(t)(1-p_{\rm
d})\Theta(t)+\frac{k}{\langle k\rangle} (1-p_{\rm
d})(1-\rho_k(t))\rho(t)[1+p_{\rm b}\Theta(\rho(t))]. \label{rateeqk}
\end{equation}
\end{widetext}
Imposing stationarity $\partial_t \rho_k(t) =0$, we obtain
\begin{equation}
  \rho_k=\frac{\frac{k}{\langle k\rangle} (1-p_{\rm
d})\rho(1+p_{\rm b}\Theta(\rho))}{\cosh+\frac{k}{\langle k\rangle}
(1-p_{\rm d})\rho(1+p_{\rm b}\Theta(\rho))}.
  \label{rhok}
\end{equation}
This set of equations imply that the higher the node connectivity,
the higher the probability to be in an occupied state. This
inhomogeneity must be taken into account in the computation of
$\Theta(\rho)$. Multiplied Eq.~(\ref{rateeqk}) by $P(k)$ and summing
over $k$, we obtain a rate equation for $\rho(t)$
\begin{equation}
   \frac{d \rho(t)}{d t} = -\rho(t)p_{\rm
d}+p_{\rm b}(1-p_{\rm d})[1-\Theta(\rho(t))]\rho(t) \Theta(\rho(t)),
   \label{rateeq}
\end{equation}
Notable, the above equation is consistent with Eq.~(\ref{rateeq00})
by imposing that $\Theta(\rho(t))=\rho(t)$ in homogeneous networks.
It also has three stationary states
\begin{equation} \label{sf steady states}
 \bar\rho^{SF}_0 = 0, \quad \Theta(\bar\rho^{SF}_\pm)=\bar\rho_\pm=\frac{1}{2}\left(1\pm\sqrt{1-\frac{4p_{\rm         d}}{p_{\rm b}(1-p_{\rm d})}}\right).
\end{equation}
The critical death rate $p_{{\rm d}_{\rm c}} =p_{\rm b}/(4+p_{\rm
b})$ is the same as that in homogeneous networks. Imposing,
naturally, $\frac{d \Theta(\rho)}{d \rho}>0$, we find that
$\bar\rho^{SF}_+$ and $\bar\rho^{SF}_0$ are stable stationary
states, whereas $\bar\rho^{SF}_-$ is unstable and therefore
represents a critical density $\rho^{SF}_{\rm c}$ below which
extinction will occur in all cases. In SF networks, the higher the
node connectivity, the higher the probability to be in an occupied
state (Eq.~(\ref{rhok})). The presence of nodes with very large
degree results in that $\bar\rho^{SF}_\pm <
\Theta(\bar\rho^{SF}_\pm)$. We conclude that both the population
steady state and the critical population in SF networks are smaller
than those in homogeneous networks.

\begin{figure}
\includegraphics[width=\columnwidth]{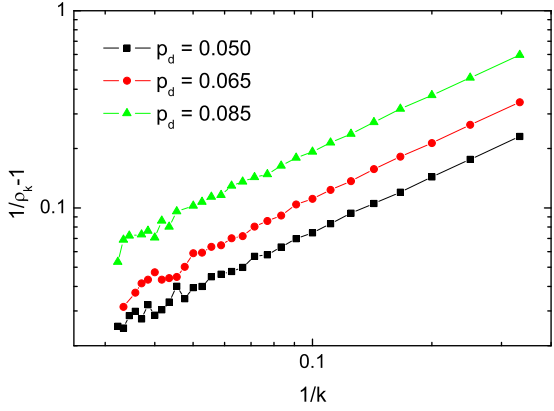}

\caption{Simulation results of $\frac{1}{\rho_{k}}-1$ against the
reciprocal of $k$ in log-log scale on uncorrelated random SF
networks with exponent $\gamma=3.0$, $k_{min}=3$ and $N=10^{3}$. The
birth rate is $p_{\rm b}=0.5$. All the plots recover the form
predicted in Eq.~(\ref{rhok}).}\label{rhokk}
\end{figure}

\begin{figure}
\includegraphics[width=\columnwidth]{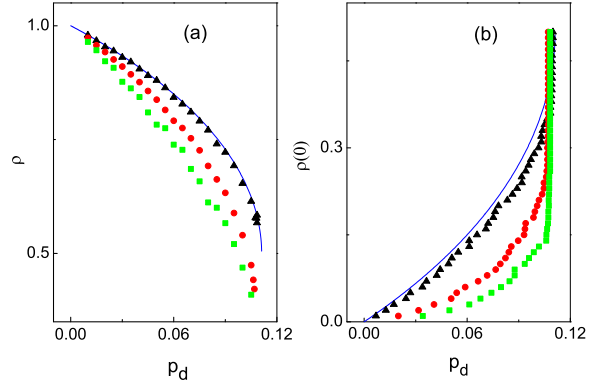}
\caption{ (a) Stationary population densities $\rho$ against the
death rate $p_{\rm d}$. (b) For different death rate $p_{\rm d}$,
there are different critical initial population densities
$\rho(0)$. All the data were obtained with $p_{\rm b}=0.5$ and
$N=1000$. MF results for the homogenous network (line) and
simulation results for the ER random network (triangles),
uncorrelated random SF networks with exponent $\gamma=3.0$
(circles) and $2.5$ (squares).}\label{rhopd}
\end{figure}

The numerical simulations performed on uncorrelated SF networks
confirm the picture extracted from the analytic treatment. We
construct the uncorrelated SF network by the algorithm
\cite{ucmmodel} with minimum degree $k_\text{{min}}=3$ and size
$N=1000$. The simulations are carried out on an initially fully
occupied network for obtaining $\rho_k$ and steady states. To find
the critical population density, due to its instability, we instead
use the initial population density $\rho (0)$ and find the value of
$p_{\rm d}$ that separates the active and absorbing states. The
prevalence $\rho_{k}$ is computed and averaged over $100$ times for
each network configuration, which are performed on $10$ different
realizations of the network. Figure \ref{rhokk} shows the behavior
of the probability $\rho_{k}$ that a node with degree $k$ is
occupied. The numerical value of the slope in log-log scale is about
$0.98$, which is in good agreement with the theoretical value $1$ in
Eq. (\ref{rhok}). In Fig. \ref{rhopd}, for ER networks with average
degree $\langle k\rangle=14$ and size $N=1000$, both plots of the
stationary population density $\rho$ and the critical initial
population density $\rho(0)$ are consistent with the MF results in
homogenous networks (Eq. (\ref{steady states})). For SF networks,
the population and the critical initial population density in steady
states are smaller than that of homogeneous networks, which agrees
with Eq. (\ref{sf steady states}). Furthermore, the more
heterogeneous the SF network, i.e., the smaller degree exponent of
the SF network, the smaller the densities. Noting that in despite of
the different network topologies, the critical death rate $p_{{\rm
d}_{\rm c}}$ is changeless, which is the prediction of Eq. (\ref{sf
steady states}).

\begin{figure}
\includegraphics[width=\columnwidth]{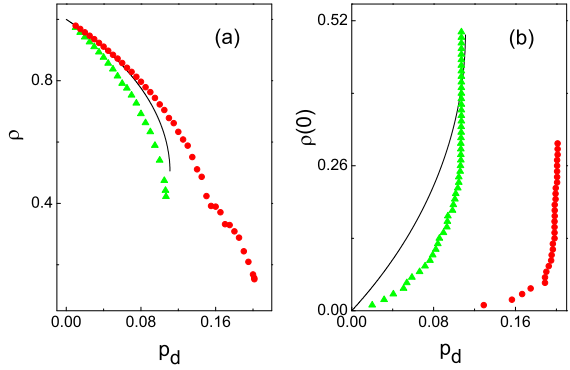}

\caption{(a) Stationary population densities $\rho$ against the
death rate $p_{\rm d}$. (b) Critical initial population densities
$\rho(0)$ versus the death rate $p_{\rm d}$. All the data were
obtained with $p_{\rm b}=0.5$ and $N=1000$. MF results for the
homogenous network (line), simulation results in the uncorrelated
random SF networks with exponent $\gamma=3.0$ with fully random
diffusion (triangles), and the modified diffusion under $\alpha=1$
and $\beta=-2$ (circles).}\label{modd}
\end{figure}

In the above model, the neighboring node is chosen with full
randomness, and we call this fully random diffusion. In the
following, we shall redesign the diffusion strategies of the
population model on SF networks. If the randomly chosen particle
does not die, a nearest neighbor node $j$ is randomly chosen with a
probability proportional to $k^{\alpha}_j$. If the neighboring node
$j$ is empty, the particle moves there; otherwise, the particle
reproduces with probability $p_{\rm b}$ producing a new particle on
another neighboring node $l$ which is chosen by a probability
proportional to $k^{\beta}_l$, conditional on that node being empty.
We can also write the MF equation in the uncorrelated random SF
network for this modified diffusion case
\begin{equation}
   \frac{d \rho(t)}{d t} = -\rho(t)p_{\rm
d}+p_{\rm b}(1-p_{\rm d})(1-\Theta_{\beta}(\rho(t)))\rho(t)
\Theta_{\alpha}(\rho(t)),
   \label{rateeqab}
\end{equation}
where
\begin{equation}
  \Theta_{\alpha}(\rho(t)) = \frac{1}{\langle
k^{1+\alpha}\rangle} \sum_k k^{1+\alpha} P(k) \rho_k(t).
  \label{eq:citaab}
\end{equation}
In the extinction state $\rho(t)=0$, it is natural that
$\Theta_{\alpha}(\rho(t))=0$. We can obtain at lowest order in
$\rho(t)$, $\Theta_{\alpha}(\rho(t))\simeq A\rho(t)$. Similarly
$\Theta_{\beta}(\rho(t))\simeq B\rho(t)$ can also be obtained at
lowest order in $\rho(t)$, where $A$ and $B$ are coefficients. In
the steady state, following previous analysis, we can easily get the
critical death rate
\begin{equation}
p_{{\rm d}_{\rm c}} =\frac{p_{\rm b}}{(4\frac{B}{A}+p_{\rm b})},
\label{pdc}
\end{equation}
which can be changed by the ratio $B/A$. If $B<A$, the modified
model has larger critical death rates than the original. Similar to
Eq. (\ref{rhok}), the partial density takes the form
$(\frac{1}{\rho_k}-1)\sim 1/D(k)$ with
$D(k)=\frac{k^{1+\alpha}}{\langle k^{1+\alpha} \rangle}+
\frac{k^{1+\beta}}{\langle k^{1+\beta} \rangle}p_{\rm
b}\Theta_{\alpha}$. Considering that $0<p_{\rm b}$ and
$\Theta_{\alpha}<1$, we can negate the second term in $D(k)$. If
$\alpha>-1$, the higher degree node has larger partial density
$\rho_k$. As $\beta<\alpha$, we get $B<A$; On the other hand, if
$\alpha<-1$, the higher degree node has smaller $\rho_k$. As $\beta
> \alpha$, we have $B<A$.

In Fig. \ref{modd} the diffusion coefficients are $\alpha=1$ and
$\beta=-2$. From the previous discussion, we obtain $B<A$. Thus, the
modified model has a larger critical death rate. Furthermore, it has
the larger steady population density and the smaller critical
population density. From the view of conservation, it is better that
a population system has a larger critical death rate, a larger
steady population density and a lower critical population density at
the same time. Our modified model has this nice property under the
condition of $B<A$.

In summary, we have studied a simple RD population model on SF
networks by analytical methods and computer simulations. We find
that in the case of fully random diffusion, the network topology can
not change the value of the critical death rate $p_{{\rm d}_{\rm
c}}$. However, the more heterogenous the network, the smaller steady
population density and the critical population density. For the
modified diffusion strategy, we can obtain the larger critical death
rate and the higher steady population density, at the meanwhile, the
lower critical population density, which is good for the specie's
survival. In present work, we consider the population model which
has only one specie, it may be more interesting to investigate
population models having multi-species with predator-prey,
mutualistic, or competitive interactions in complex networks, which
is left to future work.

This work was partially supported by NSFC/10775060, SOCIALNETS, and
POCTI/FIS/61665.

\newpage

\end{document}